\documentclass{ws-ijmpe}

\usepackage{graphicx}
\usepackage[latin1]{inputenc}

\newcommand{\AM}{$A\,$MeV}
\newcommand{\NZcp}{(N/Z)$_{\mathrm{CP}}$}
\newcommand{\NZqp}{(N/Z)$_{\mathrm{QP}}$}
\newcommand{\Ed}{$\mathrm{E_{diss}/E_{c.m.}}$}

\begin{document}

\markboth{\footnotesize E. Galichet et al.}%
{\footnotesize Isospin diffusion in $^{58}$Ni-induced reactions at intermediate energies}

\catchline{}{}{}{}{}

\title{
ISOSPIN DIFFUSION IN $^{58}$Ni-INDUCED REACTIONS AT INTERMEDIATE ENERGIES
}

\author{\footnotesize E. GALICHET\footnote{and Conservatoire National des
Arts et Métiers, F-75141 Paris, France}, \underline{M. F. RIVET}, B. BORDERIE
}

\address{Institut de Physique Nucléaire, IN2P3/CNRS, 
Université Paris-Sud-11, F-91406 Orsay, France \\
rivet@ipno.in2p3.fr}

\author{M. COLONNA}

\address{Laboratori Nazionali del Sud, I-95123 Catania, Italy}

\author{R. BOUGAULT, D. DURAND, N. LE NEINDRE, O. LOPEZ, L. MANDUCI, E. VIENT}

\address{LPC, ENSICAEN et Université, IN2P3/CNRS, F-14050 Caen, France}

\author{A. CHBIHI, J. D. FRANKLAND, J. P. WIELECZKO}

\address{GANIL, CEA et IN2P3/CNRS, F-14076 Caen, France}

\author{R. DAYRAS, C. VOLANT}

\address{IRFU/SPhN, CEA Saclay, F-91191, Gif-sur-Yvette, France}

\author{D. C. R. GUINET, P. LAUTESSE}

\address{IPN et Université Claude 
Bernard LyonI, IN2P3/CNRS, F-69622 Villeurbanne, France}

\author{M. PARLOG}

\address{NIPNE, 
RO-76900 Bucharest-M\v{a}gurele, Romania}

\author{E. ROSATO, M. VIGILANTE}

\address{Dipartimento di Scienze Fisiche e Sezione INFN, Universit\`a
Federico II, I-80126 Napoli, Italy}
\author{INDRA Collaboration}

\maketitle

\begin{history}
\received{(received date)}
\revised{(revised date)}
\end{history}

\begin{abstract}
Isospin diffusion is probed as a function of the
dissipated energy by studying two systems $^{58}$Ni+$^{58}$Ni and
$^{58}$Ni+$^{197}$Au, over the incident energy 
range 52-74\AM. Experimental data are compared with the results of a
microscopic transport model with two different parameterizations of the
symmetry energy term. A better overall agreement between data and
simulations is obtained when using a symmetry term with a potential part
linearly increasing with nuclear density. The isospin equilibration 
time at 52~\AM{} is estimated to 130$\pm$10~fm/$c$.
\end{abstract}

\section{Introduction}
Collisions between nuclei with different charge asymmetries may carry 
important information on the structure of the nuclear equation of state 
(EOS) symmetry term in density and temperature regions away from the 
normal value, that may be encountered along the reaction
path. For instance, the symmetry energy 
behaviour influences reaction processes, such as fragmentation, 
pre-equilibrium emission, N/Z equilibration between the two collisional 
partners.~\cite{Bar05,Bao08} 
Among the sensitive observables, in semi-peripheral collisions, one can 
study isospin diffusion, either through isospin transport
ratios~\cite{Tsa04,Tsa09} or by using
the isotopic content of light particle and IMF emission and  the 
asymmetry (N/Z) of the
reconstructed quasi-projectiles (QP) and quasi-targets (QT).
\cite{She04,I71-Gal09,Gal09}
The degree of equilibration, that is related to the interplay between the
reaction time and the typical time for isospin transport,
can give information about important transport properties, such as
drift and diffusion coefficients, and their relation with the 
density dependence of the symmetry energy.   

 We have performed this kind of investigations by studying 
isospin transport effects on the reaction dynamics 
for two systems, with the same projectile, $^{58}$Ni, 
and two different targets,$^{58}$Ni and $^{197}$Au, 
at incident energies of 52\AM{} and 74\AM{}.\cite{I71-Gal09,Gal09}
The N/Z ratio of the two composite systems is N/Z=1.07 for Ni+Ni 
and N/Z=1.38 for Ni+Au. 
This choice gives access to isospin effects in different conditions 
of charge (and mass)
asymmetry and to \emph{their evolution with the energy deposited 
into the system}.   
In the symmetric Ni + Ni system  isospin effects are essentially due to the 
pre-equilibrium emission.
On the contrary, in the charge (and mass) asymmetric reactions, 
one can observe isospin transport between the two partners.
The dependence of these mechanisms on the symmetry energy behaviour is
discussed.

\section{Experiment}
\subsection{Experimental details}
$^{58}$Ni (179 $\mu$g/cm$^2$) and $^{197}$Au (200 $\mu$g/cm$^2$)
targets were bombarded with  $^{58}$Ni projectiles accelerated to 52 and 
74\AM{} by the GANIL facility. The charged products emitted in collisions 
were collected by the $4\pi$ detection array INDRA.~\cite{I3-Pou95}
All elements were identified within one charge unit up to the projectile 
charge. Moreover H to Be isotopes were separated when their energy was 
high enough (above 3, 6, 8~MeV for p, d, t; 20-25~MeV for He isotopes; 
$\sim$60~MeV for Li and $\sim$80~MeV for Be).
In the following we shall call fragments the products for which only the
atomic number is measured (Z$\geq$5).
As the on-line trigger required four fired modules of the array, 
the off-line analysis only considered events in which four charged 
products were identified.~\cite{I71-Gal09}

\subsection{Event selection}

After eliminating events in which less than 90\% of the charge of 
the projectile was detected, we build subevents containing the
 ensemble of charged products which have a velocity higher than 
half the laboratory projectile velocity (i.e. forward emitted in
the nucleon-nucleon (NN) frame).
We call this ensemble ``quasi-projectile'' (QP)  
without prejudice on the equilibration of any degree of freedom 
of the system so defined. This choice eliminates the detection biases 
on the quasi-target (QT)
side and avoids considering products coming from the QT. Finally to get
 homogeneous samples we ask that the total charge of the QP is comprised 
 between 24 and 32.
In all cases 1.3 to 2$\times 10^6$ events are kept,
amounting to 14-18\% of the reaction cross sections.~\cite{I71-Gal09}

\subsection{The sorting variable}
\begin{figure}[htbp]
\centering
\includegraphics[scale=0.5]{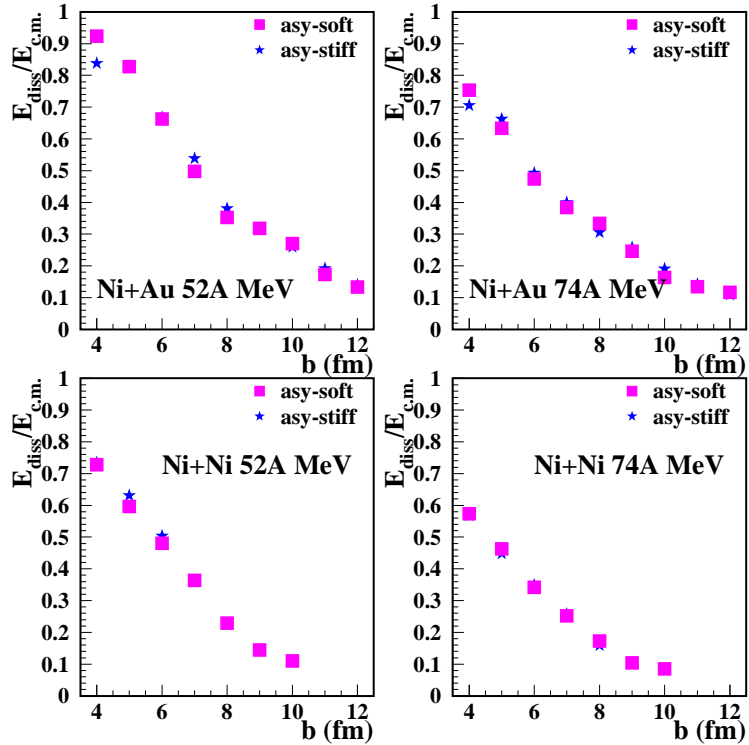}%
\caption{Correlation between \Ed{} and the impact parameter
for the four studied systems and two asy-EOS. 
From.~\protect\cite{Gal09}} \label{bEdiss}
\end{figure}
To follow the isospin diffusion as a function of the violence of the
collision, we chose to sort the events as a function
 of the dissipated energy, calculated in a binary hypothesis without mass
transfer:
\begin{equation} \label{eq:Eexc}
\mathrm{E_{diss}=E_{c.m.}}-\frac{1}{2}\mu \mathrm{V_{rel}}^2 ,
\end{equation}
with $\mu$ the initial reduced mass, $\mathrm{E_{c.m.}}$ the available 
energy and
\begin{equation} \label{eq:vrel}
\mathrm{V_{rel}=V_{QP}^{rec}} \times
\frac{\mathrm{A_{tot}}}{\mathrm{A_{target}}}
\end{equation} 
$\mathrm{V_{QP}^{rec}}$ is the QP velocity, reconstructed from the 
velocity of all the fragments it contains. Most of the QP events 
have only one fragment. For the Ni+Ni system,
about 25-30\% of the events have two or more fragments 
while for the Ni+Au system this percentage is smaller ($\sim$15\%).
It is demonstrated in~\cite{Yan03,Pia06} that the velocity of the QP is 
a good parameter for following dissipation. The dissipated energy
was calculated following the same procedure in the BNV simulations 
(see next section), and it
is shown in fig~\ref{bEdiss} that there is a close correlation between 
\Ed{} and the impact parameter. A complete description of the experimental
findings can be found in.~\cite{I71-Gal09}

\section{The BNV transport model}
We follow the reaction dynamics solving the BNV transport equation, 
that describes the evolution of the one-body distribution function 
according to the nuclear mean-field and including the effects of
two-body collisions.  
The test-particle prescription is adopted, using the TWINGO 
code~\cite{Gua96} with 50 test particles per nucleon.
The main ingredients that enter this equation are the nuclear matter 
compressibility, the symmetry energy term and its density dependence
and the nucleon-nucleon cross section.  
We take a soft isoscalar equation of state, with a compressibility
modulus K =200 MeV, which is favoured e.g. from flow studies or from the
confrontation data-dynamical simulations at intermediate 
energies.~\cite{Dan02,Bor08}

Two different 
prescriptions for the behaviour of the symmetry energy are used
to study the sensitivity of the results to the considered parameterization.
Indeed one can write the symmetry energy part of the EOS as the sum of a
kinetic term and a potential term (to be multiplied by $[(N-Z)/A]^2$), 
often approximated, for the sake of comparison, as: 
\begin{equation} \label{Esym}
\frac{E_{sym}}{A}(\rho)=\frac{C_{s,k}}{2} 
 (\frac{\rho}{\rho_0})^{2/3}+ 
\frac{C_{s,p}}{2} (\frac{\rho}{\rho _0})^{\gamma} 
\end{equation}
where $\rho_0$ is the nuclear saturation density.
The value of the $\gamma$ exponent, valid close to $\rho_0$,
determines whether the equation is 
``asy-stiff''($\gamma \geq$1, potential term continuously increasing with
$\rho$) or ``asy-soft'' ($\gamma <$1, potential term presenting a maximum 
between $\rho_0$ and 2$\rho_0$). 
In this paper we chose as asy-stiff a  potential symmetry term
linearly increasing with nuclear density ($\gamma$=1), and as asy-soft 
 the $SKM^*$ parameterization in which the potential symmetry term 
 can be approximated with $\gamma$=0.5.~\cite{Bar02} 
 
 The free
nucleon-nucleon cross section with its angular, energy and isospin
dependence was used. Indeed, as there is no consistency between the mean
field and the residual interaction in transport codes, there is no 
reason to adopt an in-medium correction which may be ``valid'' with 
one specific mean field. For instance good reproductions of nuclear 
dynamics above 
40\AM{} were obtained using the free nucleon-nucleon cross 
section.~\cite{Had96,I43-Gal03}

For the two reactions, we have ran different impact 
parameters, from b=4~fm to b=10~fm for the Ni+Ni system
and from b=4~fm to b=12~fm for the Ni+Au system. 
For each impact parameter
 10 events were produced (one event represents already the mean 
trajectory of the reaction), for the two cases of symmetry energy 
parameterization. The detailed analysis of the results of the simulations
can be found in.~\cite{Gal09}

\section{Isospin diffusion}

The measure of isospin diffusion ideally requires the determination of
the ratio N/Z for hot quasi-projectiles, just after QP and QT separate.
Experimentally it is however not possible to fully reconstruct the 
primary QP because neutrons are not measured and the masses of the 
heavier fragments are not known. 
We chose to construct an isospin-dependent variable from the isotopically
identified particles included in the QP:  
\begin{equation}
\mathrm{(<N>/<Z>)_{CP}} = \sum_{N_{evts}}{\sum_{\nu} {N_{\nu}}}
/ \sum_{N_{evts}}{\sum_{\nu} {P_{\nu}}}
\end{equation}
where $N_{\nu}$ and $P_{\nu}$ are respectively the numbers of neutrons and
protons bound in particle $\nu$ , $\nu$ being d, t, $^3$He, $^4$He, $^6$He,
$^6$Li, $^7$Li, $^8$Li, $^9$Li, $^7$Be, $^9$Be, $^{10}$Be; free protons are
excluded.  $N_{evts}$ is the number of events contained in the dissipated
energy bin considered. 
The variable \NZcp{} was calculated twice: first considering
particles forward emitted in the nucleon-nucleon frame
($\mathrm{V_{particle}>V_{proj}^{lab}}/2$), and secondly keeping
only particles forward emitted in the QP frame 
($\mathrm{V_{particle}>V_{QP}^{rec}}$). Indeed as the BNV simulation
does not allow to identify mid-rapidity particles and light fragments, 
the theoretical value of \NZcp{} is calculated only from the  particles 
evaporated by the QP, which we experimentally approximate as the particles
forward emitted in the QP frame. 

\subsection{Isospin diffusion in BNV simulations}
\begin{figure}[htbp]
\centering
\includegraphics[scale=0.5]{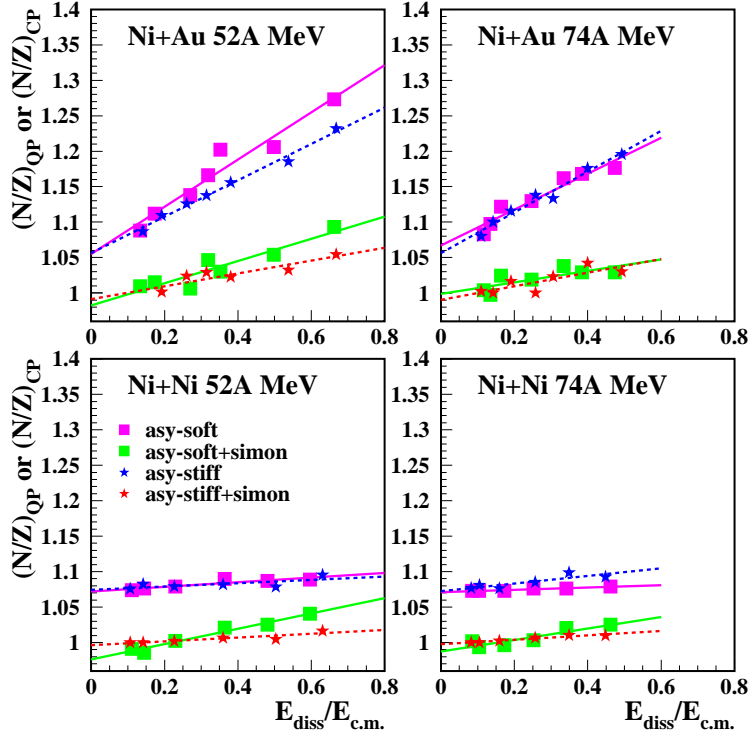}
\caption{Upper curves in each panel: isospin ratio of hot QP, 
\NZqp{}, vs \Ed{} obtained in the simulation with two asy-EOS 
(blue stars for asy-stiff and pink squares for asy-soft). The evolution of
the variable \NZcp{} obtained after de-exciting the QP with SIMON is
represented by the lower curves (red stars and green squares). 
The lines correspond to linear fits. Adapted from.~\protect\cite{Gal09}
} \label{N/ZBNV}
\end{figure}
The evolution of the N/Z ratio of the quasi-projectiles  as a function 
of \Ed{} is reported in fig.~\ref{N/ZBNV}. 
\NZqp{}  increases with the centrality of the collision
for the two systems and the two beam energies. 
For the Ni+Ni system the variation of \NZqp{} with centrality is 
small, and attributed to pre-equilibrium emission. Little dependence 
on the EOS appears at 52~\AM{}, while \NZqp{} grows slightly higher 
at 74~\AM{}  for the asy-stiff case.
Indeed, for this system with a small neutron excess, more protons are 
emitted during the pre-equilibrium stage due to the coupled effect of
Coulomb repulsion and a less attractive symmetry potential for protons. 
This effect increases with the incident energy. On the contrary,
the asy-soft case tends to emit more pre-equilibrium neutrons leading
to a lower N/Z ratio.~\cite{Lio05}

The evolution with centrality is much more pronounced for the neutron-rich
and asymmetric Ni+Au system. In addition to pre-equilibrium effects,  
isospin transport takes place between the two partners of the collision, 
and increases with the violence of the collision. 
\NZqp{} is mostly higher in the asy-soft  than in the asy-stiff case  
 for the two energies. The largest value reached, corresponding to
 b=4~fm, is lower  at 74~\AM{} than at 52~\AM{};
this may be attributed to the shorter reaction times, and to the fact that
the collision becomes more transparent. 
Thus the N/Z diffusion 
appears related to the degree of dissipation reached in 
the system and to the driving force provided by the symmetry term
of the nuclear EOS, that speeds up the isospin equilibration among 
the reaction partners.
An asy-soft EOS, more dissipative, favours isospin equilibration 
between the two partners, as found also in other recent theoretical 
investigations.~\cite{Shi03,Tsa04,Bar05,Che05,Riz08}

\subsection{Effects of secondary decay}
In figure~\ref{N/ZBNV} are also plotted
the results concerning the variable \NZcp{}, calculated after de-exciting
the hot primary QPs with the help of the SIMON code.~\cite{Dur92}
The values of \NZcp{} are always smaller and the evolution 
with dissipation is generally flatter  than that of the N/Z of 
the primary QP: \emph{secondary decay weakens the isospin effects}.

At 52~\AM{} however the differences between the results of the two 
parameterizations are more pronounced for \NZcp{}, with respect 
to \NZqp{}. Indeed excitation energies are larger in the asy-soft case,
which favours the emission of neutron-richer particles, and thus 
enhances the effect due to the larger N/Z value observed
in that case for the QP. At 74~\AM{}, no discrimination between the 
asy-EOS is possible for Ni+Au (as was already the case for hot QPs),
whereas for the reaction Ni+Ni  
 the effects due to the de-excitation modify the
initial trend imposed by the dynamical evolution: the values of 
\NZqp{} are larger in the asy-stiff case; conversely the associated
\NZcp{}  are smaller than those obtained in the asy-soft case.

Finally \NZcp{}, which appears in the simulations to be linearly
correlated with the N/Z of the primary QP, is a good indicator of 
isospin transport effects and is sensitive to the asy-EOS.

\subsection{Comparison between experimental and simulated data}

\begin{figure}[htbp]
\centering
\includegraphics[scale=0.5]{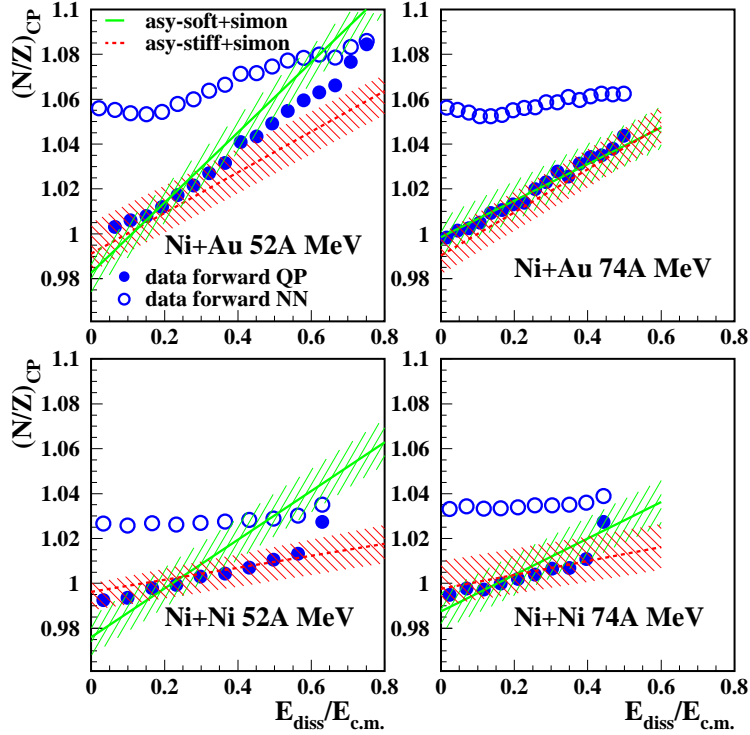}
\caption{Isospin ratio of complex particles, \NZcp{}, 
vs \Ed{}. Circles correspond to experimental data: open,
forward in NN frame, close, forward in QP frame. Error bars 
are within the size of the symbols. Dotted and solid lines are
as in fig~\ref{N/ZBNV} and hatched zones give error bars from
simulations.
Adapted from.~\protect\cite{Gal09}
} \label{N/Zcp}
\end{figure}

The evolution of the experimental values of \NZcp{} with dissipation
is displayed in fig~\ref{N/Zcp}. 
Open points show the values obtained forward 
in the NN frame. Let us remind that in this case we mix mid-rapidity
particles and those coming from the QP de-excitation.
For the Ni+Ni system at both incident energies, \NZcp{}
varies by at most 1.5\% when dissipation increases. This is the expected
behaviour for this symmetric system where N/Z is only modified by
pre-equilibrium emission, as explained in the previous subsection.
For the Ni+Au system the isospin ratio is higher
than that of the Ni+Ni system whatever the dissipated energy.
For the three first bins of 
dissipated energy  \NZcp{} has the same value for the two incident
energies and slightly decreases with dissipation; this is due 
to the on-line trigger.~\cite{I71-Gal09}
The isospin ratio of the Ni+Au system is however higher than 
that of the Ni+Ni system, which could arise from
the neutron skin of the Au target and/or from the mid-rapidity
particles included in our quasi-projectile selection which are more neutron 
rich.~\cite{I23-Lef00,I17-Pla99} This result is a first indication of 
isospin diffusion.
At higher dissipated energies,  \NZcp{} presents a significant increase
with dissipation and reaches higher values at 52\AM, 
while the trend is flatter at 74\AM.
This may be interpreted as a progressive isospin diffusion when 
collisions become more central, in connection with a larger overlap 
of the reaction partners and thus a longer interaction time. For a 
given centrality, the separation time is longer at 52\AM{} than at 
74\AM, leaving more time to the two main partners to go towards 
isospin equilibration.

The close points in fig~\ref{N/Zcp} are related to the values of
\NZcp{} forward in the QP frame. They are in all cases smaller than the
previous ones, and for Ni+Au at both energies, they grow faster 
with dissipation. This is because the mid-rapidity particles are no 
longer included: it is known that these particles are more 
neutron-rich, and  that their isospin content is independent of the 
violence of the collision.~\cite{I23-Lef00}
The values of \NZcp{} forward in the QP frame are compared with the 
results of the simulation, displayed in fig~\ref{N/Zcp} by the lines 
and the hatched zones. 
A first result  worth mentioning is that the chemical composition (N/Z) 
of the quasi-projectile forward emission appears as a very good 
representation of the composition of the entire quasi-projectile source. 
Such an observation seems to validate a posteriori the
selection frequently used to characterize the QP de-excitation 
properties.

When looking \emph{globally} at the results for the four
cases treated here, the agreement is better when the asy-stiff EOS is used,
i.e. a linear increase of the potential term of the symmetry energy around
normal density.  Note however that for Ni+Au at 52~\AM{}, where isospin
transport effects are dominant, the close points lie in between the 
simulated results  with the two EOS. This observation allows us to put an
error bar on our result, expressed as $\gamma$=1$\pm$0.2 
(see Eq. (\ref{Esym})). This value
is in reasonable agreement with that recently obtained from isospin
diffusion in Sn+Sn systems at 50\AM{}~\cite{Tsa09}, 
and matches  the one derived from the competition
between dissipative mechanisms for Ca+Ca,Ti at 25 \AM{}.~\cite{Amo09}

\section{Isospin equilibration time}

\begin{figure}[htbp]
\begin{minipage}[c]{0.475\textwidth}
\includegraphics[scale=0.5]{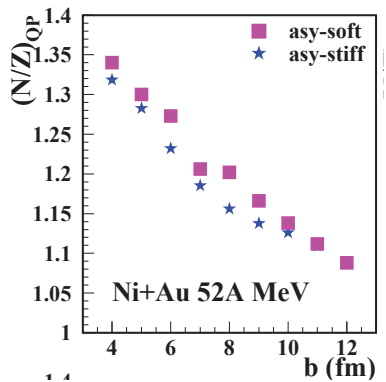}%
\end{minipage}%
\hspace*{0.05\textwidth}%
\begin{minipage}[c]{0.475\textwidth}
\caption{Isospin ratio of primary Ni QP 
vs the impact parameter obtained in BNV simulations 
for the Ni+Au system at 52~\AM{}.
Adapted from.~\protect\cite{Gal09}} \label{NZqp}
\end{minipage}%
\end{figure}
The knowledge of the different time scales associated to the various 
degrees of freedom involved in heavy-ion collisions at intermediate 
energy is of crucial importance to determine the physical properties 
of nuclear sources produced in the exit channel. Particularly
with the announced 
exotic beams the N/Z degree of freedom will hopefully be explored over 
a wide range, and thus an estimate of the chemical (isospin) equilibration 
time becomes essential. It turns out that we can extract this information
from the present data. Fig~\ref{NZqp} shows, for the Ni+Au system at 
52~\AM{}, the evolution of the N/Z of the primary QP with the 
impact parameter obtained in BNV simulations. The
equilibrium value, N/Z=1.38, is almost reached for b=4~fm, at the time when
the QP and QT separate. This time is equal to 130$\pm$10~fm/$c$.
A similar value of 115~fm/$c$ has been reported  for the Zn+Au
system at 47\AM{}.~\cite{Kow07}
Interestingly enough the equilibration of isospin can be derived directly
from the experimental data. In the top-left panel of fig~\ref{N/Zcp}
one observes that the open and close points superimpose at high dissipation:
 the values of \NZcp{} are the same at mid-rapidity and at
velocities close to that of the QP. This is a strong indication of
isospin equilibration. Note that more than 75\% of the energy must have been
dissipated before equilibrium is reached. Qualitatively  this high value 
of the dissipated energy at isospin equilibration pleads for an
asy-stiff EOS, as was shown when considering isospin transport
ratios in simulated Sn+Sn collisions at 35 and 50\AM{}.~\cite{Riz08}

\section{Conclusions}
 We studied isospin transport as a function of dissipation or
centrality in heavy-ion collisions for two beam energies, 
looking directly at the average isotopic content of the light particles
emitted by the quasi-projectile.

Experimentally we followed an isospin-dependent variable, \NZcp{},
constructed with the identified isotopes belonging to quasi-projectiles, 
as a function of the dissipated energy. Its value very slightly 
increases with centrality for the Ni+Ni system, for which no isospin
transport effect is expected,  whereas it evolves much more for Ni+Au, 
indicating the presence of isospin transport.   

Simulations show that for the Ni + Ni system, the N/Z of the 
quasi-projectile is essentially determined by proton rich 
pre-equilibrium emission, and thus slightly increases with centrality. 
The effect is more pronounced using an asy-stiff equation of state.
For the Ni+Au system isospin transport takes place
and the N/Z is larger in the asy-soft case. After de-excitation of 
the hot QPs, the isospin effects, although weaker, are still present
and the variable \NZcp{} is sensitive to the asy-EOS. The non-negligible
effect of evaporation necessitates  a precise knowledge of the
evaporation properties of medium-mass exotic nuclei; the INDRA 
Collaboration has initiated an experimental program in that aim.

The results of the simulations are analysed in the same way as 
the experimental data. 
This demonstrates that \NZcp{} is linearly related to \NZqp{}.
When considering the four cases under study we find a better overall 
 agreement between simulations and
experimental data for the asy-stiff case corresponding to a 
potential symmetry term linearly increasing with nuclear density. 
Referring to Eq. (\ref{Esym}) we propose a value $\gamma$=1$\pm$0.2.
Information
concerning the isospin equilibration time,
is also obtained. At 52~$A\,$MeV for the Ni+Au and the most
dissipative collisions we can infer from the data-model comparison that
isospin equilibration is reached at 130 $\pm$ 10 fm/c. 


\end{document}